 \documentclass[pmlr,twocolumn,10pt]{jmlr} 

\usepackage{booktabs}
\usepackage{siunitx}
\usepackage{afterpage}
\usepackage{verbatim}
\usepackage{multirow}
\usepackage{breqn}
\usepackage{pdflscape}
\usepackage{appendix}
\usepackage{graphicx}
\usepackage{float}
\usepackage{colortbl}
\usepackage{makecell}

\theorembodyfont{\upshape}
\theoremheaderfont{\scshape}
\theorempostheader{:}
\theoremsep{\newline}

\jmlrvolume{LEAVE UNSET}
\jmlryear{2023}
\jmlrsubmitted{LEAVE UNSET}
\jmlrpublished{LEAVE UNSET}
\jmlrworkshop{Machine Learning for Health (ML4H) 2023} 

\title[Swin UNETR++]{Swin UNETR++: Advancing Transformer-Based Dense Dose Prediction Towards Fully Automated Radiation Oncology Treatments}

\author{%
\Name{Kuancheng Wang}\Email{kwang601@gatech.edu}\\
\addr Georgia Institute of Technology, Atlanta, GA, USA%
\AND
\Name{Hai Siong Tan}\Email{haisiong.tan@pennmedicine.upenn.edu}\\
\addr Department of Radiation Oncology, Perelman School of Medicine, University of Pennsylvania, Philadelphia, PA, USA%
\AND
\Name{Rafe Mcbeth}\Email{Rafe.Mcbeth@pennmedicine.upenn.edu}\\
\addr Department of Radiation Oncology, Perelman School of Medicine, University of Pennsylvania, Philadelphia, PA, USA%
}

\begin{document}

\maketitle

\begin{abstract}
The field of Radiation Oncology is uniquely positioned to benefit from the use of artificial intelligence to fully automate the creation of radiation treatment plans for cancer therapy. 
This time-consuming and specialized task combines patient imaging with organ and tumor segmentation to generate a 3D radiation dose distribution to meet clinical treatment goals, similar to voxel-level dense prediction. 
In this work, we propose Swin UNETR++, that contains a lightweight 3D Dual Cross-Attention (DCA) module to capture the intra and inter-volume relationships of each patient's unique anatomy, which fully convolutional neural networks lack.
Our model was trained, validated, and tested on the Open Knowledge-Based Planning dataset. 
In addition to metrics of Dose Score $\overline{S_{\text{Dose}}}$ and DVH Score $\overline{S_{\text{DVH}}}$ that quantitatively measure the difference between the predicted and ground-truth 3D radiation dose distribution, we propose the qualitative metrics of average volume-wise acceptance rate $\overline{R_{\text{VA}}}$ and average patient-wise clinical acceptance rate $\overline{R_{\text{PA}}}$ to assess the clinical reliability of the predictions. 
Swin UNETR++ demonstrates near-state-of-the-art performance on validation and test dataset (validation: $\overline{S_{\text{DVH}}}$=1.492 Gy, $\overline{S_{\text{Dose}}}$=2.649 Gy, $\overline{R_{\text{VA}}}$=88.58\%, $\overline{R_{\text{PA}}}$=100.0\%; test: $\overline{S_{\text{DVH}}}$=1.634 Gy, $\overline{S_{\text{Dose}}}$=2.757 Gy, $\overline{R_{\text{VA}}}$=90.50\%, $\overline{R_{\text{PA}}}$=98.0\%), establishing a basis for future studies to translate 3D dose predictions into a deliverable treatment plan, facilitating full automation.
\end{abstract}


\section{Introduction}
\label{sec:intro}
Radiotherapy remains a critical tool in the fight against cancer, employing sophisticated techniques like intensity-modulated radiation therapy (IMRT) for precise radiation delivery. 
Yet, even as artificial intelligence (AI) has emerged to assist with organ segmentation in the clinic, the development of patient-specific treatment plans remains a highly complicated and time-consuming task that lies outside the reach of current AI-based approaches. 
This complexity originates from trade-off decisions that must balance the maximum radiation dose to the tumor while minimizing the dose to surrounding normal tissues.
Furthermore, converting physician treatment plan intent to deliverable linear accelerator or ``robot" sequenced motions involves solving a complex inverse optimization problem using proprietary simulated annealing algorithms within dedicated treatment planning systems. 
To address this problem, several groups have worked to develop techniques to translate anatomical information, i.e., 3D images and associated organ segmentation, to a predicted radiation dose distribution that matches clinical constraints \citep{C3D, nguyen2019feasibility}.
Competitions such as Open Knowledge-Based Planning Grand Challenge (OpenKBP) \citep{babier2021openkbp} have been initiated to challenge the community to make progress on this problem, but current approaches still depend on fully convolutional models that may not be sufficient for generating deliverable treatment plans due to the lack of large receptive fields. 
The goal of this work is to transition fully convolutional 3D dose prediction to attention-based.

We introduced Swin UNETR++, a novel deep learning architecture that aims to supplant the extant fully convolutional neural network (FCNN) approaches in 3D dose prediction. 
This new model is backboned by Swin UNETR \citep{hatamizadeh2021swin} and incorporates a lightweight 3D Dual Cross-Attention (DCA) module designed to bridge the semantic gap between encoder and decoder features in the skip connection. 
This architecture excels at capturing both channel and spatial dependencies, thereby providing a more holistic view of the patient’s anatomy for accurate dose prediction.
Swin UNETR++ was rigorously evaluated using the OpenKBP dataset and trained using a hybrid of mean squared error loss $\mathcal{L}_{\text{MSE}}$ and a novel global differentiable approximation of the dose-volume histogram loss $\mathcal{L}_{\text{DVH\_global}}$. 
In terms of performance metrics, Swin UNETR++ shows near-state-of-the-art results, outpacing most existing models in both the validation and test phases, and its implementation marks a significant advancement in automating the radiotherapy planning process. 
This study not only outlines the architecture and performance of Swin UNETR++ but also opens avenues for further research in incorporating more nuanced attention mechanisms into treatment planning.
\section{Related Work}
After the OpenKBP dataset was released, several teams have shown the strength of full CNN. \cite{nguyen20193d} proposed Hierarchically Densely Connected U-net (HD U-net). \cite{soomro2021deepdosenet} introduced DeepDoseNet, which integrates Resnet \citep{he2016deep} and dilated densenet blocks in their architecture. \cite{C3D} proposed cascade 3D model, C3D, which ranked No.1 on OpenKBP leaderboard for both $\overline{S_{\text{DVH}}}$ and $\overline{S_{\text{Dose}}}$. In C3D, two U-nets are connected in series, and the second U-net refines the coarse prediction made by the first U-net. As ViT \citep{dosovitskiy2020image} and Swin UNETR result in state-of-the-art performance in classification and segmentation tasks, we use Swin UNETR++ to demonstrate the superiority of the transformer-based approach in 3D radiation dose estimation. 
\section{Method}
\label{sec:math}

\subsection{Dataset}
The OpenKBP dataset contains 340 patients treated by IMRT for head-and-neck cancer. Each patient was prescribed a combination of three possible radiation doses: 70 Gy to the gross disease planning target volume (PTV70), 63 Gy to the intermediate-risk planning target volumes (PTV63), and 56 Gy to the elective planning target volumes (PTV56), with organs-at-risk (OARs) masks including the brain stem, spinal cord, right parotid, left parotid, esophagus, larynx, and mandible. In addition, CT, the ground-truth 3D radiation dose distribution, and a possible dose mask contouring the body are provided for each patient. All image data are downsampled to 128 × 128 × 128 voxels with voxel dimensions around 3.5 mm × 3.5 mm × 2 mm. In terms of dataset distribution, OpenKBP Grand Challenge assigned patients No.1--200 for training, No.201--240 for validation, and No.241--340 for testing.

\subsection{Model Architecture}

\begin{figure*}[h!]
\floatconts
  {fig:arch1}
  {\caption{Swin UNETR++ model architecture highlighting 3D dual cross-attention (DCA) module}}
  {\includegraphics[width=1\textwidth]{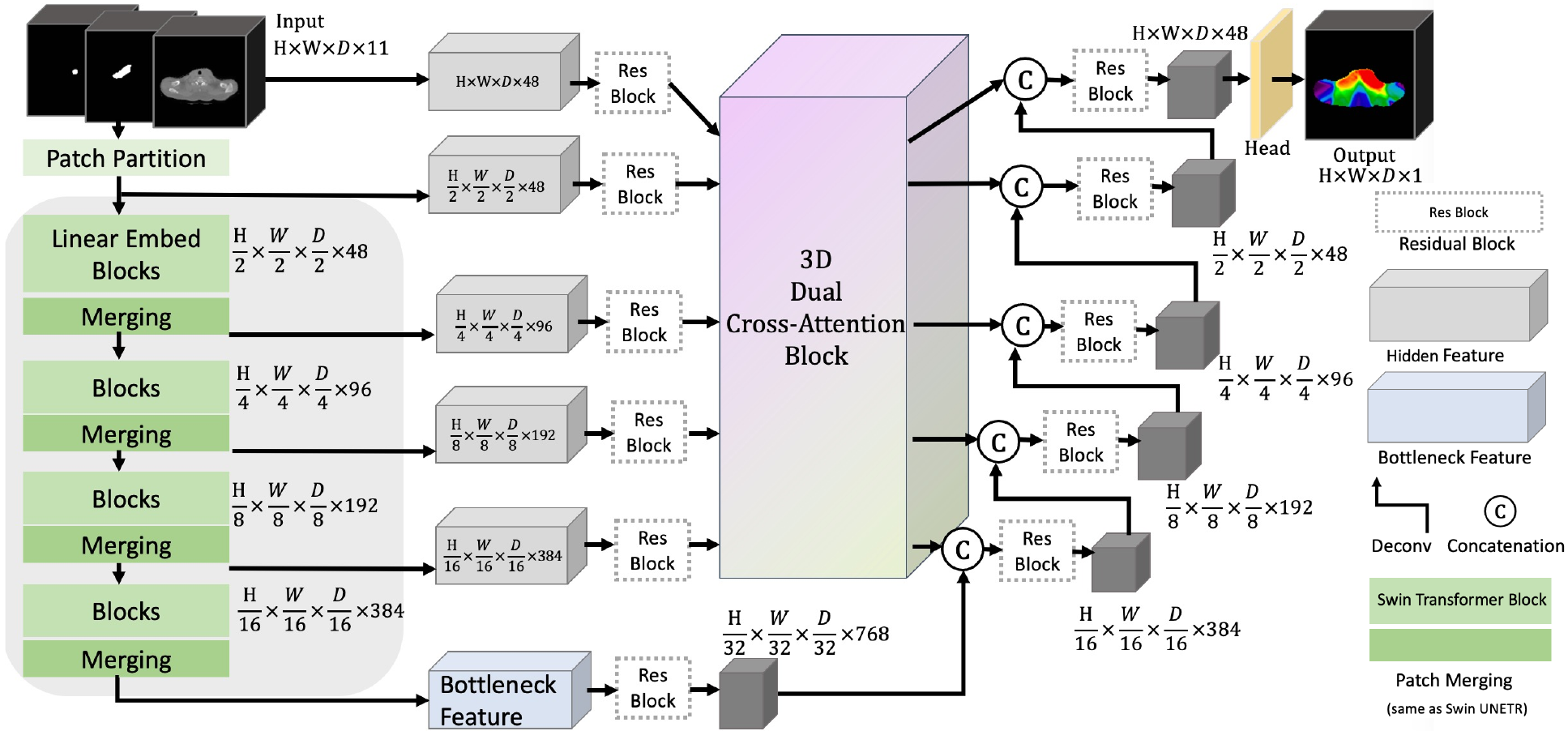}}
\end{figure*}

\begin{figure*}[h!]
 \floatconts
  {fig:arch2}
  {\caption{DCA module internal structure}}
  {\includegraphics[width=\textwidth]{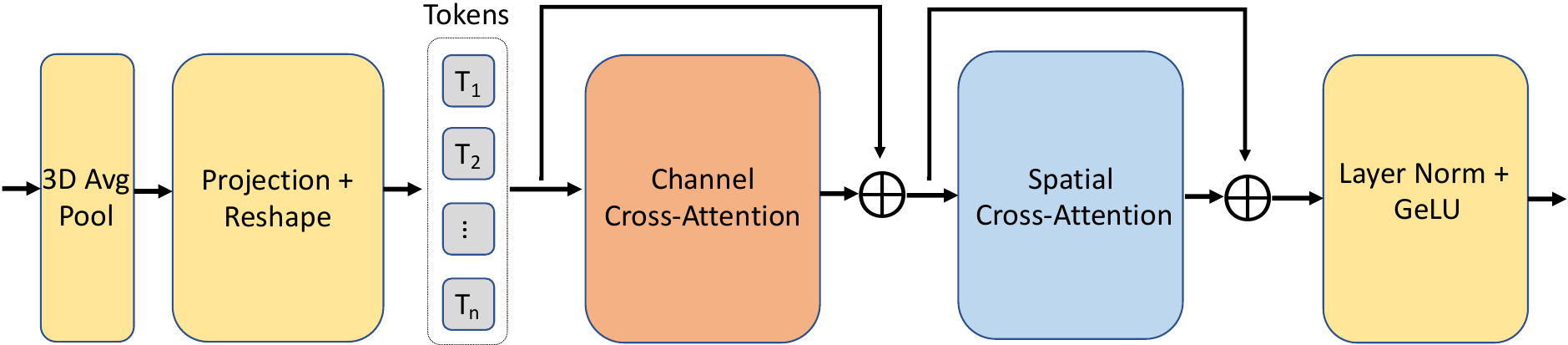}}
\end{figure*}
   
\subsubsection{Swin UNETR}
\label{sec:op}
Swin UNETR has recently emerged as a promising method for medical image segmentation challenges, registering state-of-the-art performance in specialized domains such as glioma delineation, Medical Segmentation Decathlon \citep{simpson2019large}, and Beyond The Cranial Vault Segmentation Challenge \citep{tang2022self}. 
The model combines the computational advantages of the Swin Transformer \citep{liu2021swin}, known for its shifted window attention mechanisms with the domain-specific robustness of UNETR \citep{hatamizadeh2022unetr}.
This combination benefits from spatial hierarchy representation but also adheres to computational efficiency guidelines. 
However, despite these merits, our analysis has identified an overlooked shortcoming—namely, the semantic gap between the encoder and decoder features propagated through the skip connections. 
The discrepancies between low-level features and high-level semantics in U-Net-based architectures are corroborated by \cite{pang2019towards} and \cite{ wang2022deep}. 
Resolving this semantic inconsistency could improve the accuracy of 3D radiation dose prediction, which requires more details and clinical understanding than semantic segmentation.

\subsubsection{Swin UNETR++}
\label{sec:op}
Research has been looking at using attention mechanisms to improve the performance of U-Net-based architectures. 
For instance, \cite{MOU2021101874} introduced CS$^2$-Net, which utilized spatial and channel attention to facilitate the segmentation of curvilinear structures. 
The DANet \citep{fu2019dual} integrated position and channel attention module in scene segmentation. 
Recently, \cite{ates2023dual} proposed a light-weight 2D plug-and-play Dual Cross-Attention (DCA) module which has demonstrated its efficiency in reducing the semantic gap among common fully convolutional U-Net-based architecture.

In extending the conventional 2D DCA module to 3D, the fundamental operations, such as convolutional layers, pooling operations, and batch normalization, have been extended from 2D to their 3D counterparts. 
It comprises three essential stages: multi-scale 3D patch embedding, a 3D Channel Cross-Attention (CCA) module, and a 3D Spatial Cross-Attention (SCA) module. 

\textbf{3D Patch Embedding.} We used the same patch size defined by \cite{ates2023dual}. Given $n$ encoder stages and encoder features $E_i \in \mathbb{R}^{C_i\times \frac{H}{2^{i-1}}\times \frac{W}{2^{i-1}}\times \frac{D}{2^{i-1}}}$ where $i\in \{1,2,...,n\}$, the flattened patches for stage $i$ are extracted by 3D $1\times 1 \times 1$ depth-wise convolutions and 3D average pooling layer as following: 
  \begin{equation}
  T_{i}= \text{DConv1D}_{E_i}(\text{Reshape}(\text{AvgPool3D}_{E_i}(E_i)))
  \end{equation}
  
\textbf{
3D Channel Cross-Attention (CCA) module.} We concatenated $T_i$ along the channel dimension to create $T_c$. By using the 3D $1\times 1 \times 1$ depth-wise convolutions, we can obtain the queries $Q_i$, projected keys $K$, values $V$, and CCA as follows:
  \begin{equation}
  Q_i=\text{DConv1D}_{Q_i}(T_i)
  \end{equation}
  \begin{equation}
  K=\text{DConv1D}_{K}(T_c)
  \end{equation}
  \begin{equation}
  V=\text{DConv1D}_{V}(T_c)
  \end{equation}
  \begin{equation}
      C_c=\text{total num of channels}
  \end{equation}
  \begin{equation}
  \text{CCA}(Q_i,K,V)=\text{Softmax}\left(\frac{Q_{i}^{T}K}{\sqrt{C_c}}\right)V^T
  \end{equation}

\textbf{3D Spatial Cross-Attention (SCA) module.} With the reshaped output from CCA $\Bar{T_i}$, we can obtain $\Bar{T_c}$ by layer normalization and channel concatenation. Similarly, we can obtain the queries $Q_i$, keys $K$, values $V$, and SCA as follows:
  \begin{equation}
  Q=\text{DConv1D}_{Q}(\Bar{T_c})
  \end{equation}
  \begin{equation}
  K=\text{DConv1D}_{K}(\Bar{T_c})
  \end{equation}
  \begin{equation}
  V_i=\text{DConv1D}_{V_i}(\Bar{T_i})
  \end{equation}
  \begin{equation}
  h=\text{num of heads for CCA}
  \end{equation}
  \begin{equation}
  \text{SCA}(Q,K,V_i)=\text{Softmax}\left(\frac{QK^{T}}{\sqrt{\frac{C_c}{h}}}\right)V_i
  \end{equation}

By combining the 3D DCA module in Figure~\ref{fig:arch2} with the Swin UNETR architecture in the skip connection, we proposed Swin UNETER++ illustrated in Figure~\ref{fig:arch1}. It improves medical image analysis and is specifically tailored for dense radiation dose prediction. 
It is also the first time the 3D DCA module was applied outside semantic segmentation. 

The 3D DCA module has clinical importance as well. CCA refines the model's understanding of the channel-wise relationship between different clinically significant volumes, similar to the prior work of overlap volume histogram \citep{wu2009patient} that quantified the distance between different OARs and PTVs.
SCA establishes the close-and-distant relationships within each clinically significant volume. 
Together, CCA and SCA aim to heighten treatment efficacy while minimizing radiation-induced complications, contributing to a safer, more effective therapeutic regimen.

\subsection{Network Training}
\subsubsection{Data Preprocessing and Augmentation}
Inspired by \cite{C3D}, we first clipped the CT images into a fixed range between -1024 and 1500 HU and then normalized them by 1000 HU during preprocessing. Any missing volume masks for each patient are represented by a 3D zero tensor.

During training, we applied data augmentation with MONAI \citep{cardoso2022monai} to increase our model's generalizability and avoid overfitting in the following order: (1)Random axis flip which randomly flips the spatial axis of CT, structured masks, and possible dose mask (probability=0.5). (2)Random affine which performs random shearing in a range of $(0.1, 0.1, 0.1)$ with zero-padding and nearest-neighbor interpolation on CT, structured masks, and possible dose mask (probability=0.2). (3)Random zoom scales CT, structured masks, and possible dose mask within a range of $[0.9, 1.3]$ with nearest-neighbor interpolation (probability=0.5). (4)Random Guassian smooth is applied only to CT (probability=0.2). (5)Random contrast is applied only to CT (probability=0.2).
 
\subsubsection{Loss Function}
To compare the voxel difference between the predicted radiation dose distribution $D_{\text{pred}}$ and the ground-truth radiation dose distribution $D_{\text{gt}}$,  the mean squared error (MSE) was used due to its sensitivity to outliers. 
Therefore, we designed the loss function $\mathcal{L}_{\text{MSE}}$, which only focuses on the dose difference within the patient's body, using the possible dose mask $M_d$. For all possible voxel indices $i, j, k$:
\begin{align}
&\mathcal{L}_{\text{MSE}}(D_{\text{pred}},D_{\text{gt}})\\ 
\nonumber
&=\frac{\sum_{i,j,k} (D_{\text{pred}}(i, j, k) - D_{\text{gt}}(i, j, k))^2 M_{\text{d}}(i, j, k)}{\sum_{i,j,k} M_{\text{d}}(i, j, k)}
\end{align}


The DVH loss $\mathcal{L}_{\text{DVH}}$ \citep{nguyen2020incorporating} is a loss function that takes into account a common clinical tool for assessing the dose to each organ in a treatment plan, the dose volume histogram (DVH). As the traditional DVH calculation is not differentiable, a differentiable approximate dose-volume histogram $\widetilde{\text{DVH}}$ is used.
Given a set of $n_v$ clinically significant volume masks $M_v$, where $v\in V=\text{PTV}\cup\text{OAR}$,  and the number of bins $n_t$, we can obtain:
\begin{align}
&\mathcal{L}_{\text{DVH}}(D_{\text{pred}}, D_{\text{gt}}, M_v) \\
\nonumber
&= \frac{\sum_{v \in V} \| \widetilde{\text{DVH}}_v(D_{\text{gt}}, M_v) - \widetilde{\text{DVH}}_v(D_{\text{pred}}, M_v) \|_2^2}{n_vn_t}
\end{align}

We introduced a variant of the DVH loss, the global DVH loss $\mathcal{L}_{\text{DVH\_global}}$.
The global DVH condenses the integral dose across the patient, i.e., the possible dose mask segmentation $M_d$, to a single line on the DVH plot.
The loss then considers the difference of this integral dose distribution between the predicted and ground truth distributions.
This loss considers the radiation dose across the entire patient's body as opposed to individual clinically segmented volumes that are sparse and account for only a fraction of the total patient volume. $\mathcal{L}_{\text{DVH\_global}}$ is computed as follows:
\begin{align}
&\mathcal{L}_{\text{DVH\_global}}(D_{\text{pred}}, D_{\text{gt}}, M_d) \\
\nonumber
&= \frac{\| \widetilde{\text{DVH}}_v(D_{\text{gt}}, M_d) - \widetilde{\text{DVH}}_v(D_{\text{pred}}, M_d) \|_2^2}{n_t}
\end{align}

We combined $\mathcal{L}_{MSE}$ and $\mathcal{L}_{\text{DVH\_global}}$ together as the final loss function for training :
\begin{equation}
\mathcal{L}_{\text{combined}}=\alpha \mathcal{L}_{MSE}+\beta\mathcal{L}_{DVH\_global}
\end{equation}
where both $\alpha$ and $\beta$ were set to 10 empirically.

\subsubsection{Hyperparameter Setting}
Adam optimizer \citep{kingma2014adam} was initialized with a learning rate of 2e-4 and beta values of (0.5, 0.999).
To adaptively adjust the learning rate during the training process, we utilized the Cosine Annealing with Warm Restarts scheduler. 
The initial restart period was set to 100 epochs, with a multiplicative factor of 1 and a minimum learning rate of 2e-5. Furthermore, we fixed the seed to 42, thereby ensuring deterministic behavior for stochastic operations and facilitating experiment reproducibility.

\section{Evaluation Metrics}
\subsection{Quantitative Metrics}
\subsubsection{Dose Score}
Provided by OpenKBP, the Dose Score $\overline{S_{\text{Dose}}}$ is the mean absolute difference between the predicted radiation dose distribution $D_{\text{pred}}$ and the ground-truth radiation dose distribution $D_{\text{gt}}$ based on the possible dose mask $M_{\text{d}}$ for each patient $P_t$ in either validation or test datasets.
\begin{align}
&\text{Error}_{\text{Dose}}(P_t) \\
\nonumber
&= \frac{\sum_{i,j,k}^{} \left| D_{\text{pred}}(i,j,k) - D_{\text{gt}}(i,j,k) \right|M_{\text{d}}(i,j,k)}{\sum_{i,j,k}^{} M_{\text{d}}(i,j,k)}
\end{align}
\begin{equation}
\overline{S_{\text{Dose}}} = \frac{1}{n} \sum_{t=1}^{n} \text{Error}_{\text{Dose}}(P_t)
\end{equation}

\subsubsection{DVH Score}
DVH Score $\overline{S_{\text{DVH}}}$ \citep{babier2021openkbp} is designed to measure the quality of the predicted radiation dose distribution clinically. The radiation dose distribution for each OAR is evaluated by:\\

\noindent$\mathcal{D}_{0.1cc}$:=The minimum max dose to all but 0.1cc
$\mathcal{D}_{mean}$:=The mean dose (i.e., the area under DVH curve)\\

The radiation dose distribution for each PTV is evaluated by:\\

\noindent$\mathcal{D}_{1}$:=The minimum max dose received by 99\% of the target (i.e., 99th percentile)\\
$\mathcal{D}_{95}$:=The minimum dose received by 95\% of the target (i.e., 5th percentile)\\
$\mathcal{D}_{99}$:=The minimum dose received by 99\% of the target (i.e., 1st percentile)\\

Finally, the mean absolute difference between the predictions and the ground truth radiation dose distributions is computed based on the above metrics for each patient using their existing OAR and PTV masks.

\subsection{Qualitative Metrics}
To evaluate the clinical reliability of the model's predictions, we defined the absolute difference between the ground-truth and predicted dose distribution within 3 Gy as clinically acceptable and the absolute difference outside this range as clinically unacceptable \citep{sher2021prospective, mashayekhi2021artificial}. 

\subsubsection{Volume-wise Clinical Acceptance}
We consider the predicted radiation dose distribution for a volume to be volume-wise clinically acceptable VA if and only if the absolute dose distribution difference between the predicted and the ground-truth within the given volume is less than 3 Gy. The function $\text{VA}()$ has two possible outputs: 1 represents volume-wise clinically acceptable while 0 represents unacceptable. Given a set of validation or test patients $P$ and a set of clinically significant volumes $V$, the volume-wise clinical acceptance rate $R_{\text{VA}}$ for a volume $v \in V$ is defined as:  
\begin{equation}
R_{\text{VA}}(v) = \frac{1}{|P|} \sum_{p \in P}\text{VA}(p, v)
\end{equation}
The average volume-wise clinical acceptance rate $\overline{R_{\text{VA}}}$ for an entire validation or test dataset is defined as:
\begin{equation}
\overline{R_{\text{VA}}} = \frac{1}{|V|} \sum_{v \in V} R_{\text{VA}}(v)
\end{equation}

\subsubsection{Patient-wise Clinical Acceptance}
Clinically, medical dosimetrists try to achieve Pareto optimal plans given individual patient's unique anatomy and physicians' prioritization, but different prioritization can lead to different ``optimal" plans \citep{nguyen2019generating, kyroudi2020exploration}. To simulate these real-world plan variations, we first find the absolute difference between the predicted and ground-truth radiation dose distribution for all clinically significant volumes $V$ of each patient $p$ and then average those absolute differences. We consider a patient to be patient-wise clinically acceptable PA if and only if the average difference is less than 3 Gy. The function PA() has two possible outputs: 1 represents patient-wise clinically acceptable while 0 represents patient-wise clinically unacceptable. Given a set of all validation or test patients $P$, the average patient-wise clinical acceptance rate $\overline{R_{\text{PA}}}$ for an entire validation or test dataset is defined as:
\begin{equation}
\overline{R_{\text{PA}}} = \frac{1}{|P|} \sum_{p \in P}\text{PA}(p, V)
\end{equation}

\begin{table*}[htbp]
\centering
\caption{Swin UNETR++ vs Baseline Models in terms of $\overline{S_{\text{DVH}}}$, $\overline{S_{\text{Dose}}}$, $\overline{R_{\text{VA}}}$, and $\overline{R_{\text{PA}}}$}
\vspace{0.5em}
\label{tab:swin_unetr++}
\begin{tabular}{|c|c|c|c|c|c|}
\hline
\rule{0pt}{2.6ex}Architecture & Dataset & $\overline{S_{\text{DVH}}}$ & $\overline{S_{\text{Dose}}}$ & $\overline{R_{\text{VA}}}$ & $\overline{R_{\text{PA}}}$\\
\hline
\multirow{2}{*}{Swin UNETR++} & Validation & \cellcolor{lightgray!50}\textbf{1.492} & \cellcolor{lightgray!50}\textbf{2.649} & 88.58\% & \cellcolor{lightgray!50}\textbf{100.00\%}\\
\cline{2-6}
                              & Test & \cellcolor{lightgray!50}\textbf{1.634} & \cellcolor{lightgray!50}\textbf{2.757} & \cellcolor{lightgray!50} \textbf{90.50\%} &
                              \cellcolor{lightgray!50}\textbf{98.00\%}\\
\hline
\multirow{2}{*}{Swin UNETR}  & Validation & 1.555 & 2.667 & 89.30\% & \cellcolor{lightgray!50}\textbf{100.00\%}\\
\cline{2-6}
                             & Test & 1.773 & 2.818 & 88.94\% & 
                             97.00\%\\
\hline
\multirow{2}{*}{3D U-Net}  & Validation & 2.763 & 3.325 & 68.34\% & 82.50\%\\
\cline{2-6}
                             & Test & 2.778 & 3.408 & 70.07\% & 
                             74.00\%\\
\hline
\multirow{2}{*}{Attention U-Net}  & Validation & 2.035 & 2.868 & 86.23\% & 97.50\%\\
\cline{2-6}
                             & Test & 2.079 & 3.044 & 83.19\% & 
                             94.00\%\\
\hline
\multirow{2}{*}{SegResNet}  & Validation & 2.029 & 2.858 & 79.69\% & 95.00\%\\
\cline{2-6}
                             & Test & 1.987 & 2.941 & 84.19\% & 
                             93.00\%\\
\hline
\multirow{2}{*}{DynUNet}  & Validation & 2.252 & 3.028 & 82.45\% & 92.50\%\\
\cline{2-6}
                             & Test & 2.222 & 3.096 & 85.87\% & 
                             96.00\%\\
\hline
\multirow{2}{*}{HD U-Net}  & Validation & 1.720 & 2.835 & \cellcolor{lightgray!50}\textbf{91.28\%} & \cellcolor{lightgray!50}\textbf{100.00\%}\\
\cline{2-6}
                             & Test & 1.876 & 3.003 & 87.86\% & 
                             97.00\%\\
\hline
\multirow{2}{*}{C3D}  & Validation & 1.784 & 2.727 & 86.91\% & 95.00\%\\
\cline{2-6}
                             & Test & 1.847 & 2.879 & 87.66\% & 
                             96.00\%\\
\hline
\end{tabular}
\end{table*}

\begin{table*}[h!]
\centering
\caption{Comparison between different combinations of loss functions} 
\vspace{0.5em}
\label{tab:loss}
\begin{tabular}{|c|c|c|c|c|}
\hline
\rule{0pt}{2.6ex}Architecture & Loss Function & Dataset & $\overline{S_{\text{DVH}}}$ & $\overline{S_{\text{Dose}}}$ \\
\hline
\multirow{3}{*}{Swin UNETR++} & $\mathcal{L}_{\text{DVH\_global}}$ \& $\mathcal{L}_{\text{MSE}}$ & Test & \cellcolor{lightgray!50} \textbf{1.634} & \cellcolor{lightgray!50}\textbf{2.757} \\
\cline{2-5}
                              & $\mathcal{L}_{\text{DVH}}$ \& $\mathcal{L}_{\text{MSE}}$ & Test & 1.791 & 2.916\\
\hline
\end{tabular}
\end{table*}
\section{Results}
\label{sec:results}
We compared the Swin UNETR's performance with 3D U-Net \citep{cciccek20163d}, Attention U-Net \citep{oktay2018attention}, SegResNet \citep{myronenko20193d}, DynUNet, HD U-net \citep{nguyen20193d}, C3D \citep{C3D} and its backboned architecture Swin UNETR \citep{hatamizadeh2021swin}. 3D U-Net, Attention U-Net, SegResNet, DynUNet, HD U-net, and C3D are fully CNN-based, among which HD U-net and C3D are specifically designed for 3D radiation dose estimation. The training pipeline for all architectures is the same, except we modified the loss function for C3D to adapt its cascade mechanism. The architecture results measured based on $\overline{S_{\text{DVH}}}$, $\overline{S_{\text{Dose}}}$, $\overline{R_{\text{VA}}}$, and $\overline{R_{\text{PA}}}$ are available in Table~\ref{tab:swin_unetr++}, where the highest values of each metric in each dataset are highlighted. The details of these results are explained in sections~\ref{sec:ablation} and ~\ref{sec:clinical} from the aspect of ablation study and clinical analysis.

\subsection{Ablation Study}
\label{sec:ablation}
It's clear from Table~\ref{tab:swin_unetr++} that Swin UNETR++ outperforms all seven other architectures on $\overline{S_{\text{DVH}}}$, $\overline{S_{\text{Dose}}}$, $\overline{R_{\text{VA}}}$, and $\overline{R_{\text{PA}}}$ in both validation and test datasets. The only exception is HD U-net's validation $\overline{R_{\text{VA}}}$ is 2.7\% higher than Swin UNETR++'s. Out of 50 more teams, Swin UNETR++ should rank $1^{st}$ on $\overline{S_{\text{DVH}}}$ and $9^{th}$ on $\overline{S_{\text{Dose}}}$ in the validation dataset; ranks $12^{th}$ on $\overline{S_{\text{DVH}}}$ and $10^{th}$ on $\overline{S_{\text{Dose}}}$ in the test dataset on the leaderboard.

\subsubsection{The Effect of 3D DCA Module}
By comparing the metric scores between Swin UNETR++ and Swin UNETR in Table~\ref{tab:swin_unetr++}, Swin UNETR++ lowers $\overline{S_{\text{DVH}}}$ by 0.063 Gy and lowers $\overline{S_{\text{Dose}}}$ by 0.018 Gy in the validation dataset; lowers $\overline{S_{\text{DVH}}}$ by 0.139 Gy, $\overline{S_{\text{Dose}}}$ by 0.061 Gy, increases $\overline{R_{\text{VA}}}$ by 1.56\%, and increases $\overline{R_{\text{PA}}}$ by 1\% in the test dataset. This demonstrates the performance improvement due to the 3D DCA module.

\subsubsection{The Effect of $\mathcal{L}_{MSE}+\mathcal{L}_{DVH\_global}$}
We further examined the advantage of the global DVH loss $\mathcal{L}_{\text{DVH\_global}}$ over the traditional DVH loss $\mathcal{L}_{\text{DVH}}$, which only calculates the structural DVH different based on the PTV and OAR masks, in table~\ref{tab:loss}. These two loss functions are examined on test data since there are more variations in patients' imaging, thus allowing us to better generalize the effect of $\mathcal{L}_{\text{DVH\_global}}$. $\mathcal{L}_{\text{DVH\_global}}$ demonstrates its efficiency by lowering $\overline{S_{\text{DVH}}}$ by 0.157 Gy and lowering $\overline{S_{\text{Dose}}}$ by 0.159 Gy.

\subsection{Clinical Analysis}
\label{sec:clinical}
We further investigated the potential clinical value Swin UNETR++ has. Figure~\ref{fig:dose_viz} and~\ref{fig:difference_viz} visualize each model's dose estimation and its absolute difference from ground truth on the same slice for the representative patient No.241 in the test dataset. Swin UNETR++ has higher accuracy in the high-dose volume, PTV56, PTV63, and PTV70, among other architectures. Figure~\ref{fig:dvh_viz} illustrates the dose-volume histogram for each architecture's prediction on patient No.241. Unlike other architectures, Swin UNETR++ keeps the DVH-differences smallest among all regions of interest and never overdoses the PTVs, meaning it is comparatively safer.

The clinical reliability of the AI-generated plan is another major concern. While Table~\ref{tab:swin_unetr++} provided the averaged volume acceptance rate $\overline{R_{\text{VA}}}$ and patient acceptance rate $\overline{R_{\text{PA}}}$, more details about the radiation dose distribution of each architecture for each volume in validation and test datasets can be found in Tables~\ref{tab:volume_acceptance_1} and~\ref{tab:volume_acceptance_2}. Swin UNETR++ has a higher $\overline{R_{\text{VA}}}$ and $\overline{R_{\text{PA}}}$, highlighting its reliability. It's also worth noting that there is no single model good at dose estimation for all the volumes of interest, indicating a possible improvement in the future.

\section{Discussion}
This work represents continued efforts to bridge the gap between the technical achievements of computer science and the complexities of clinical medicine. 
Adopting AI in radiation oncology requires clinical endpoints to be at the forefront of collaborations between the different domains.
Replacing humans in clinical workflows with artificial intelligence is not appropriate from many patients' perspectives, and instead, we believe these models will augment the clinical judgment and intelligence of clinical providers.
While our results are not state-of-the-art on the specified OpenKBP grand challenge, the difference between the top ten models is very small, and each teams' training pipeline is significantly different. For example, the No.1 team C3D iterated over each training patient multiple times in each epoch while we only iterated over each training patient once. We also don't know if some teams used their own patient data to pretrain the model. Based on the experiment we did in section~\ref{sec:results}, our Swin UNETR has better performance than other models such as C3D \citep{C3D} and is still representative of a model with clinical viability.

The competition dataset also does not fully represent the data in our clinic, as it contains IMRT plans that use static angles, whereas many clinics have transitioned to the use of volumetric modulated arc therapy (VMAT) plans that represent treatment along continuous arcs.
Our clinical data represents more patients, has higher resolutions, contains more organ segmentations, represents our unique patient population, and is more variable in treatment approach than the idealized challenge data. 
We recognize that this work could benefit from additional clinical testing, but the scope of this project was limited to a summer undergraduate research fellowship. 
Further clinical training and testing are underway and will provide additional results. 

Our model, which translates patient anatomy via CT images and associated segmentations to a 3D dose distribution, is the first step in a two-step process to produce a clinically deliverable radiation treatment plan. 
The second step involves generating treatment machine path sequences, i.e., multileaf collimator leaf sequences, gantry and collimator positions, etc., from the predicted 3D dose distributions. 
This step is currently solved by proprietary inverse optimization algorithms as described in \citep{otto2008volumetric} that require time-consuming manual intervention but may be solvable with more robust deep learning architectures, as shown in \citep{heilemann2023generating}.
This provided additional motivation for exploring the techniques described in this paper for future studies. 
If realized, finding solutions to the radiation treatment planning challenge will represent a large leap forward in the speed of creation and quality of treatment plans, augmenting treatment teams' intelligence and maximizing efficiency.

\section{Conclusion}
In this paper, we introduced Swin UNETR++ which is a novel transformer-based architecture for dense radiation dose prediction. It integrates the dual cross-attention module, consisting of channel cross-attention and spatial cross-attention, into Swin UNETR's skip connection to reduce the semantic gap between encoder and decoder features and capture significant clinical relationships between PTVs and OARs. Our model is examined on OpenKBP Grand Challenge dataset and demonstrates near-state-of-the-art performance on $\overline{S_{\text{DVH}}}$ and $\overline{S_{\text{Dose}}}$. Furthermore, the investigation on the clinical safety and reliability of the prediction results confirms the potential clinical value transformer-based architecture can have. We believe that transitioning dose prediction from fully convolutional to transformer-based architectures will provide a strong foundation for the future work of translating dose prediction into deliverable treatment machine parameters, completing the full integration of AI into the clinical treatment planning process.

\section{Acknowledgment}
This work was supported by NIH grant R25 CA140116.

\clearpage
\bibliography{jmlr-sample}

\clearpage

\appendix
\onecolumn
\section{Tables}
\begin{table*}[h!]
\centering
\caption{$R_{\text{VA}}$ by Structure for Each Architecture} 
\vspace{0.5em}
\label{tab:volume_acceptance_1}
\begin{tabular}{|c|c|c|c|}
\hline
Structures & Architecture & Validation $R_{\text{VA}}$ & Test $R_{\text{VA}}$\\
\hline
\multirow{7}{*}{Brainstem} & Swin UNETR++ & 91.43\% & 92.13\% \\
\cline{2-4}
& Swin UNETR & 85.75\% & 94.38\% \\
\cline{2-4}
& 3D U-Net & 71.43\% & 83.15\% \\
\cline{2-4}
& Attention U-Net& 88.57\% & 91.01\% \\
\cline{2-4}
& SegResNet & 80.00\% & 92.13\% \\
\cline{2-4}
& DynUNet & 82.86\% & 93.26\% \\
\cline{2-4}
& HD U-net & 91.43\% & 92.13\% \\
\cline{2-4}
& C3D & 91.43\% & 94.38\% \\
\hline
\hline
\multirow{7}{*}{Spinal Cord} & Swin UNETR++ & 89.47\% & 95.45\% \\
\cline{2-4}
& Swin UNETR &84.21\% & 95.45\%\\
\cline{2-4}
& 3D U-Net & 86.84\% & 94.32\% \\
\cline{2-4}
& Attention U-Net& 92.11\%& 94.32\% \\
\cline{2-4}
& SegResNet & 92.11\% & 89.77\% \\
\cline{2-4}
& DynUNet & 86.84\% & 95.45\% \\
\cline{2-4}
& HD U-net & 94.74\% & 96.59\% \\
\cline{2-4}
& C3D & 94.74\% & 93.18\% \\
\hline
\hline
\multirow{7}{*}{Right Parotid} & Swin UNETR++ & 82.5\% & 86.87\% \\
\cline{2-4}
& Swin UNETR &92.50\% & 82.83\%\\
\cline{2-4}
& 3D U-Net & 55.00\% & 50.51\% \\
\cline{2-4}
& Attention U-Net& 92.50\% & 76.77\% \\
\cline{2-4}
& SegResNet & 77.50\% & 75.76\% \\
\cline{2-4}
& DynUNet & 82.50\% & 85.85\% \\
\cline{2-4}
& HD U-net & 90.00\% & 87.87\% \\
\cline{2-4}
& C3D & 82.50\% & 85.86\% \\
\hline
\hline
\multirow{7}{*}{Left Parotid} & Swin UNETR++ & 84.21\% & 88.78\% \\
\cline{2-4}
& Swin UNETR &78.95\% & 95.71\%\\
\cline{2-4}
& 3D U-Net & 78.94\% & 63.27\% \\
\cline{2-4}
& Attention U-Net& 81.58\% & 76.53\% \\
\cline{2-4}
& SegResNet & 78.94\% & 75.51\% \\
\cline{2-4}
& DynUNet & 73.68\% & 77.55\% \\
\cline{2-4}
& HD U-net & 94.74\% & 81.63\% \\
\cline{2-4}
& C3D & 89.47\% & 85.71\% \\
\hline
\hline
\multirow{7}{*}{Esophagus} & Swin UNETR++& 94.12\% & 79.07\% \\
\cline{2-4}
& Swin UNETR &100.00\% & 83.72\%\\
\cline{2-4}
& 3D U-Net & 88.24\% & 72.09\% \\
\cline{2-4}
& Attention U-Net& 82.35\% & 67.44\% \\
\cline{2-4}
& SegResNet & 82.35\% & 69.77\% \\
\cline{2-4}
& DynUNet & 76.47\% & 69.77\% \\
\cline{2-4}
& HD U-net & 82.35\% & 76.74\% \\
\cline{2-4}
& C3D & 88.24\% & 83.72\% \\
\hline
\end{tabular}
\end{table*}

\begin{table*}[h!]
\centering
\caption{$R_{\text{VA}}$ by Structure for Each Architecture (continued from Table~\ref{tab:volume_acceptance_1})} 
\vspace{0.5em}
\label{tab:volume_acceptance_2}
\begin{tabular}{|c|c|c|c|}
\hline
Structures & Architecture & Validation $R_{\text{VA}}$ & Test $R_{\text{VA}}$\\
\hline
\multirow{7}{*}{Larynx} & Swin UNETR++ & 77.27\% & 84.91\% \\
\cline{2-4}
& Swin UNETR &86.36\% & 73.58\%\\
\cline{2-4}
& 3D U-Net & 45.45\% & 52.83\% \\
\cline{2-4}
& Attention U-Net& 81.82\% & 75.47\% \\
\cline{2-4}
& SegResNet & 59.09\% & 75.47\% \\
\cline{2-4}
& DynUNet & 72.72\% & 81.13\% \\
\cline{2-4}
& HD U-net & 86.36\% & 77.36\% \\
\cline{2-4}
& C3D & 68.18\% & 73.58\% \\
\hline
\hline
\multirow{7}{*}{Mandible} & Swin UNETR++ & 82.14\% & 83.33\% \\
\cline{2-4}
& Swin UNETR &85.71\% & 86.11\%\\
\cline{2-4}
& 3D U-Net & 42.86\% & 54.16\% \\
\cline{2-4}
& Attention U-Net& 75.00\% & 83.33\% \\
\cline{2-4}
& SegResNet & 67.86\% & 79.17\% \\
\cline{2-4}
& DynUNet & 78.57\% & 88.89\% \\
\cline{2-4}
& HD U-net & 82.14\% & 87.50\% \\
\cline{2-4}
& C3D & 75.00\% & 83.33\% \\
\hline
\hline
\multirow{7}{*}{PTV56} & Swin UNETR++ & 97.37\% & 97.80\% \\
\cline{2-4}
& Swin UNETR &94.74\% & 96.70\%\\
\cline{2-4}
& 3D U-Net & 84.21\% & 72.53\% \\
\cline{2-4}
& Attention U-Net& 97.37\% & 96.70\% \\
\cline{2-4}
& SegResNet & 89.47\% & 97.80\% \\
\cline{2-4}
& DynUNet & 97.37\% & 97.80\% \\
\cline{2-4}
& HD U-net & 97.37\% & 96.70\% \\
\cline{2-4}
& C3D & 94.74\% & 92.31\% \\
\hline
\hline
\multirow{7}{*}{PTV63} & Swin UNETR++ & 92.31\% & 97.62\% \\
\cline{2-4}
& Swin UNETR &92.31\% & 92.86\%\\
\cline{2-4}
& 3D U-Net & 65.38\% & 61.90\% \\
\cline{2-4}
& Attention U-Net& 88.46\% & 83.33\% \\
\cline{2-4}
& SegResNet & 84.61\% & 90.48\% \\
\cline{2-4}
& DynUNet & 88.46\% & 80.95\% \\
\cline{2-4}
& HD U-net & 96.15\% & 88.10\% \\
\cline{2-4}
& C3D & 92.31\% & 90.48\% \\
\hline
\hline
\multirow{7}{*}{PTV70} & Swin UNETR++ & 95.00\% & 99.00\% \\
\cline{2-4}
& Swin UNETR &92.50\% & 98.00\%\\
\cline{2-4}
& 3D U-Net & 65.00\% & 57.00\% \\
\cline{2-4}
& Attention U-Net& 82.50\% & 87.00\% \\
\cline{2-4}
& SegResNet & 85.00\% & 96.00\% \\
\cline{2-4}
& DynUNet & 85.00\% & 88.00\% \\
\cline{2-4}
& HD U-net & 97.50\% & 94.00\% \\
\cline{2-4}
& C3D & 92.50\% & 94.00\% \\
\hline
\end{tabular}
\end{table*}

\newpage
\clearpage

\section{Result Visualizations}
\label{apd:second}
.
\begin{figure*}[htbp]
  \centering
  \caption{3D dose estimation on test patient No.241 from each architecture}
  \includegraphics[width=0.99\textwidth]{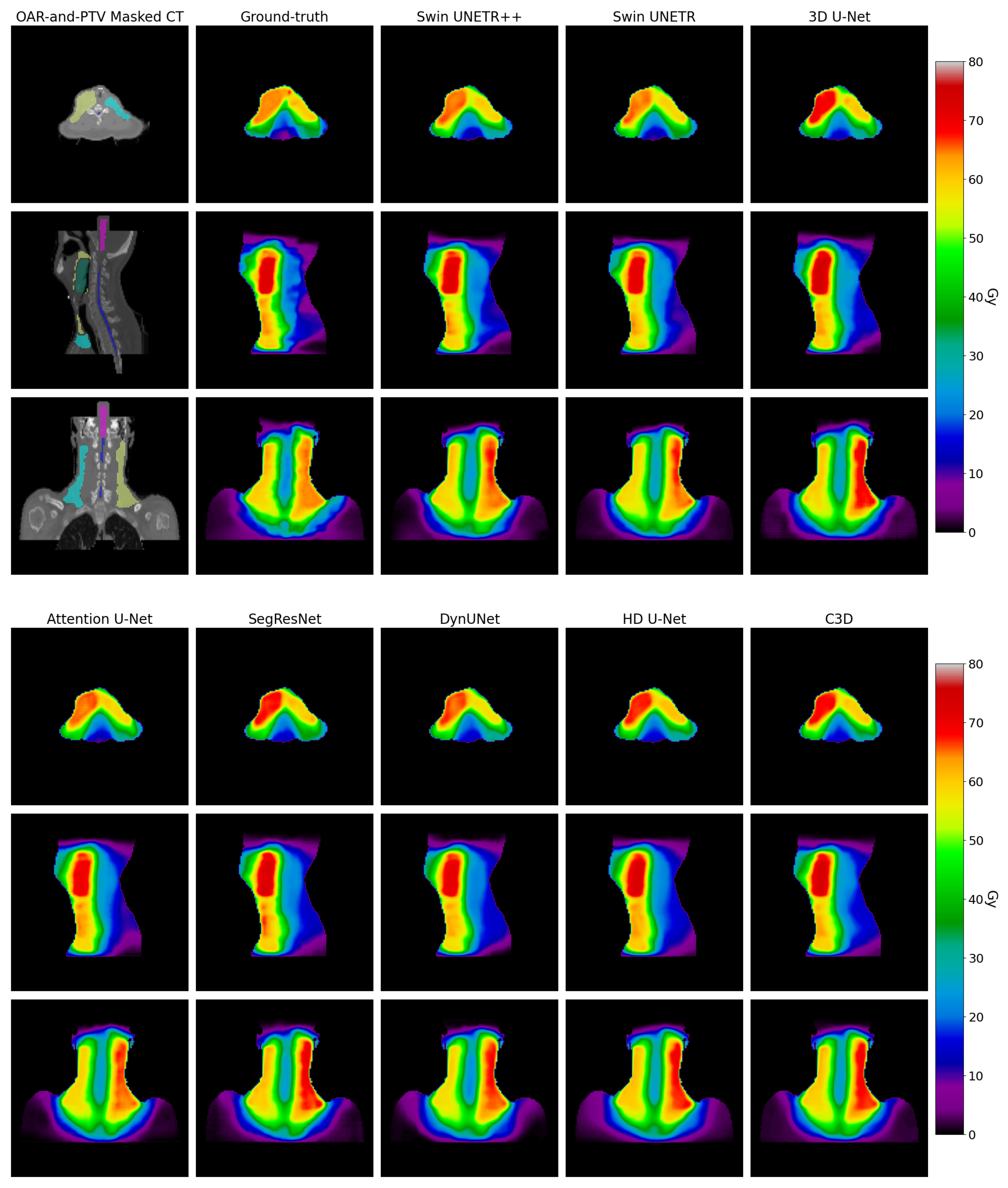}
  \label{fig:dose_viz}
\end{figure*}

\begin{figure*}[htbp]
  \centering
  \caption{$|D_{\text{pred}}-D_{\text{gt}}|$ for test patient No.241 from each architecture}
  \includegraphics[width=0.99\textwidth]{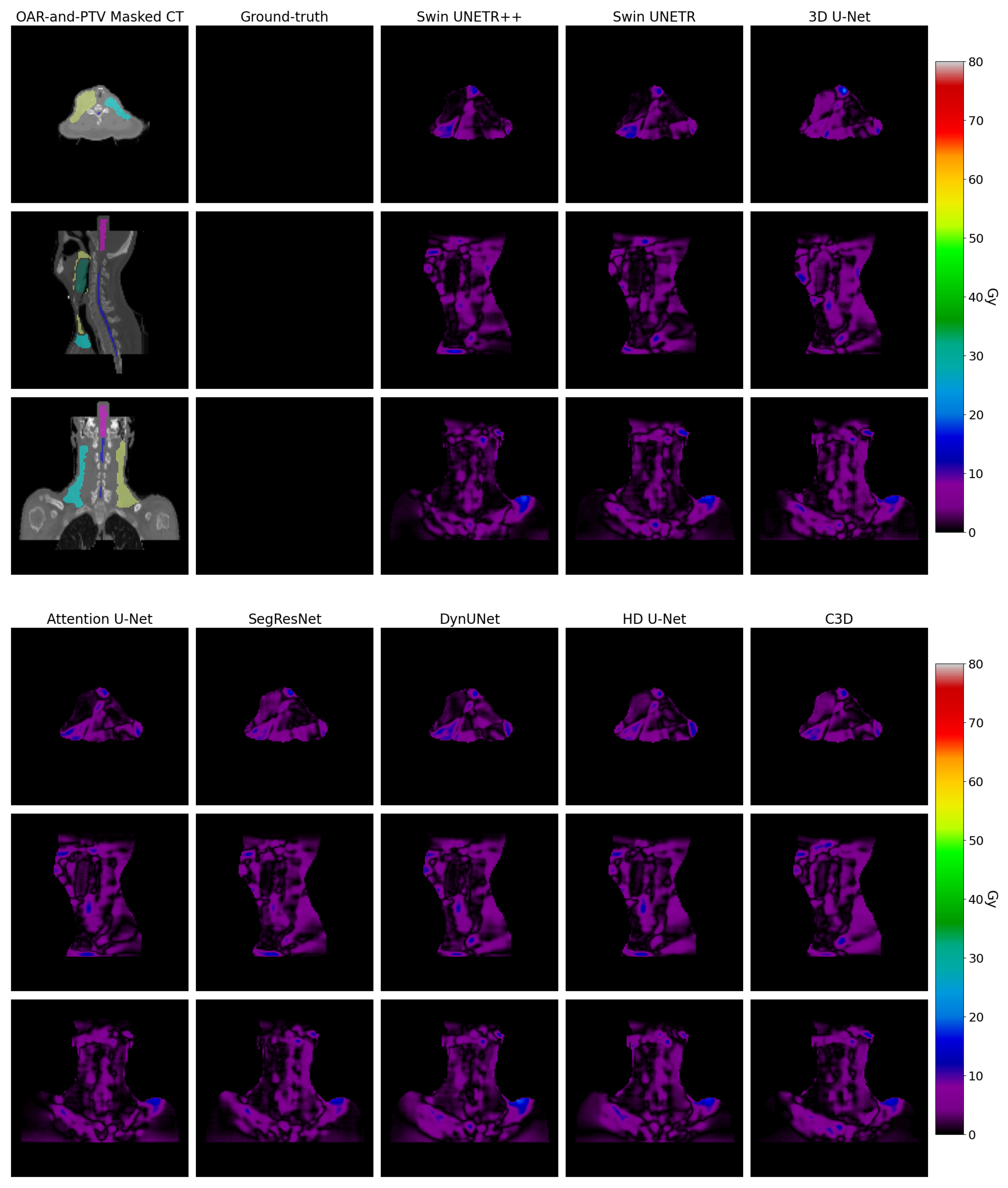}
  \label{fig:difference_viz}
\end{figure*}

\begin{figure*}[htbp]
  \centering
  \caption{DVH: predicted (dashed) vs ground-truth (solid) for test patient No.241 from each architecture}
  \includegraphics[width=0.99\textwidth]{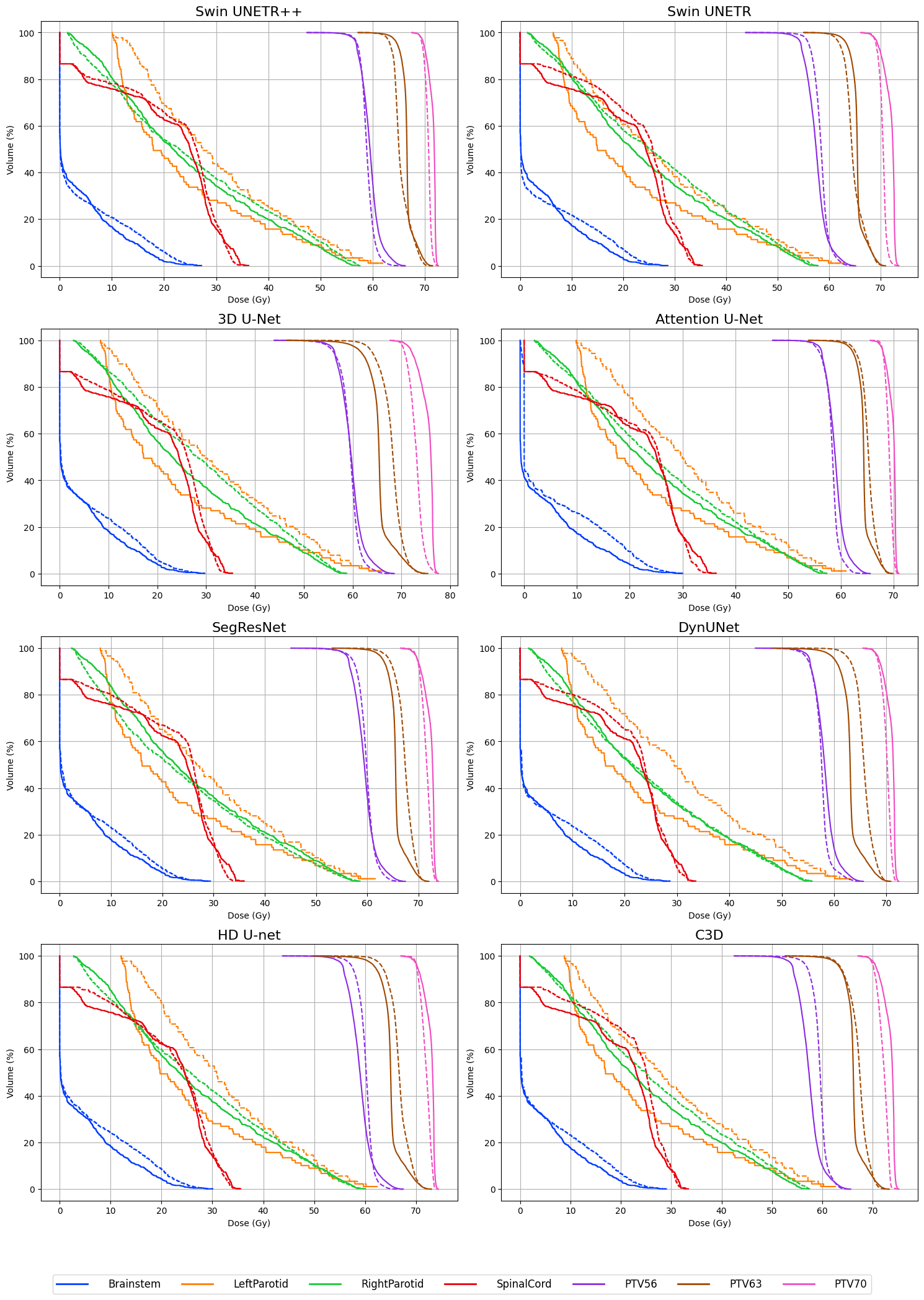}
  \label{fig:dvh_viz}
\end{figure*}
\end{document}